\documentclass[aps,pre,reprint,superscriptaddress,floatfix,amsmath,amssymb]{revtex4-2}
\usepackage{graphicx}
\usepackage{dcolumn}
\usepackage{bm}
\usepackage[usenames,dvipsnames]{xcolor}
\usepackage{hyperref}
\usepackage{placeins}
\usepackage{siunitx}

\begin{document}

\title{Characterizing and modeling the patterns of vehicle movement on road networks}

\author{Dongwon Kang}
\affiliation{Department of Physics, University of Seoul, Seoul 02504, Republic of Korea}

\author{Jung-Hoon Jung}
\affiliation{Department of Physics, University of Seoul, Seoul 02504, Republic of Korea}
\affiliation{Department of Physics, Korea University, Seoul 02841, Republic of Korea}

\author{Jongwoo Lee}
\affiliation{Department of Transport Big Data, Korea Transport Institute, 370 Sicheong-daero, Sejong 30147, Republic of Korea}

\author{Seunghoon Cheon}
\affiliation{Department of Transport Big Data, Korea Transport Institute, 370 Sicheong-daero, Sejong 30147, Republic of Korea}

\author{Young-Ho Eom}
    \email{yheom@uos.ac.kr}
\affiliation{Department of Physics, University of Seoul, Seoul 02504, Republic of Korea}
\affiliation{Urban Big data and AI Institute, University of Seoul, Seoul 02504, Republic of Korea}

\date{\today}

\begin{abstract}
Understanding vehicle movement on road networks is closely related to various practical and theoretical issues.
While recent works have focused on which cost vehicles minimize while moving, how they move to minimize that cost remains less explored.
In this work, we analyze large-scale data of individual vehicle trajectories in real-world road networks to identify cost-minimizing movement patterns of vehicles and the influence of road network structure on such movement.
We observed that vehicle movements exhibit three phases: the beginning, middle, and end of trips.
At the beginning and end, vehicles detour more, lose directional memory quickly, and travel at lower speeds than during the middle.
In contrast, during the middle, they tend to detour less, maintain directional memory, and travel faster than at the beginning and end.
Finally, at the beginning and end, vehicles exhibit similar detour and velocity patterns, except the direction of movement.
To understand these patterns, we propose a double-layered network model mimicking the hierarchical structure of real-world road networks.
We found that when vehicles move across our model network while minimizing travel time, they tend to concentrate on high-level roads, and the three observed movement phases are reproduced.
Consequently, when a vehicle moves between a given origin-destination pair, it must enter and exit these high-level roads.
This causes it to deviate from the trajectory that minimizes travel distance between the same origin-destination pair---particularly at the beginning and end of the trip.
Our results reveal common patterns underlying individual vehicle movements that appear highly diverse at first glance, demonstrating that these patterns emerge because vehicles leverage the characteristics of hierarchical road networks to minimize travel time.
\end{abstract}

\maketitle

\section{Introduction}
How do vehicles move from one location to others in real-world road networks?
This question is relevant to not only practical issues from the efficient design of road networks and urban structure~\cite{barthelemy2008ModelingUrbanStreet, clark1951UrbanPopulationDensities, louf2013ModelingPolycentricTransition, lee2017MorphologyTravelRoutes, liu2015RevealingTravelPatterns, zheng2018SpatialTemporalTravel} to traffic congestion~\cite{jung2023EmpiricalAnalysisCongestion, colak2016UnderstandingCongestedTravela, demartino2009CongestionPhenomenaComplexa} and urban air pollution~\cite{rosenlund2008ComparisonRegressionModelsa, zhang2013AirPollutionHealtha, gasana2012MotorVehicleAir}, but also theoretical issues such as the motion of active particles in network-shaped space~\cite{chowdhury2000StatisticalPhysicsVehicular, helbing2001TrafficRelatedSelfdriven, bechinger2016ActiveParticlesComplex} and human mobility in urban environments~\cite{pappalardo2015ReturnersExplorersDichotomy, tang2015UncoveringUrbanHuman, peng2012CollectiveHumanMobility}.

Current works on vehicular movement on road networks have focused on what objective function vehicles optimize when traveling between a given origin-destination (OD) pair.
In transportation engineering, a simple yet powerful idea underlying vehicle movement is formalized as Wardrop's first principle, which states that individual vehicles move to minimize their travel time under given conditions~\cite{wardrop1952RoadPaperTheoretical}.
Although slight deviations from this principle have been reported in some empirical studies, it is widely accepted that optimizing travel time is the primary objective of vehicle movement~\cite{tang2018DeviationActualShortest, wardman2004PublicTransportValues, zhu2015PeopleUseShortest, papinski2013RouteChoiceEfficiency, li2018AnalysisTaxiDrivers, yang2020PathTimeEfficiency, bekhor2006EvaluationChoiceSet}.
In network science, similarly, a pervasive idea is that flow between two nodes in a network takes place through the shortest path connecting the nodes~\cite{newman2003StructureFunctionComplex, boccaletti2006ComplexNetworksStructure, barthelemy2011SpatialNetworks, jeong2000LargescaleOrganizationMetabolic, bullmore2009ComplexBrainNetworks}, although a recent work suggested that substantial vehicular traffic flows through an alternative shortest path in some cities~\cite{akbarzadeh2018CommunicabilityGeometryCaptures}.
These works in common suggest that vehicle movement is based on minimizing some cost functions such as travel time or distance.

\begin{figure}[b]
    \centering
    \includegraphics[width=\linewidth]{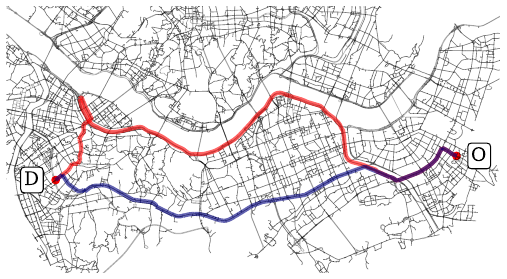}
    \caption{ Visualization of an actual vehicle trajectory (red line) and a trajectory obtained when the vehicle minimizes its travel distance for the same OD pair (blue line) on the Seoul road network.
    Note that the actual trajectory shows a significant deviation from the trajectory obtained by minimizing travel distance, though they share the same OD.
    }
    \label{fig:sample_trajectory}
\end{figure}

\begin{figure*}
    \centering
    \includegraphics[width=\linewidth]{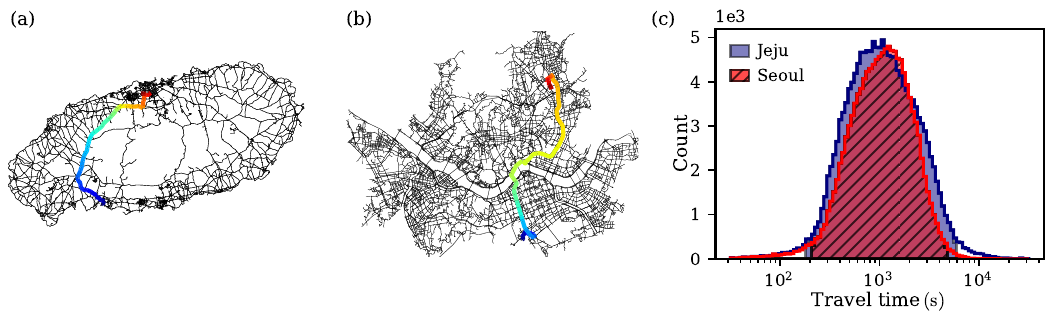}
    \caption{ Representative trajectories in (a) Jeju and (b) Seoul. For both trajectories, the starting points are marked in blue and the end points are marked in red and the colors denote the relative temporal evolution along each trajectory.
    (c) Travel time distribution of trajectories in Jeju during January 2019 (blue) and in Seoul from January 1 to January 7, 2019 (red, diagonal hatching).
    The shaded area marks trajectories within $2\sigma$ of the mean in logarithmic space, representing the dataset used for our analysis.
    Filtered travel times range from \SI{183}{\second} to \SI{5832}{\second} in Jeju and from \SI{212}{\second} to \SI{4703}{\second} in Seoul.}
    \label{fig:dataset}
\end{figure*}

However, even if we know which cost function vehicles minimize when they move, this only accounts for half of understanding their movement.
A question remains: How do vehicles move from one location to another in road networks to minimize a cost function such as travel time or distance?
More specifically, when minimizing such costs, do vehicles move persistently toward the destination with almost uniform velocity, or do they instead exhibit systematic variations in direction and speed during their movement?

Moreover, since vehicles move within a road network rather than in open space, it can be expected that the structure of the road network will play a critical role in vehicle movement.
For instance, if a road network were homogeneous---meaning vehicles could travel at the same speed on all roads---the trajectory of a vehicle passing between a given OD pair will be similar to the shortest distance path between the same OD pair, since travel time would be directly proportional to distance.
However, since real-world road networks are in general not homogeneous, the actual trajectories produced by vehicles tend to deviate significantly from the shortest distance paths between the same OD pair, as shown in Fig.~\ref{fig:sample_trajectory}.

In this work, we analyze large-scale individual vehicle trajectory data to investigate how vehicles move while minimizing their costs on real-world road networks and how they utilize the road network structure to achieve cost minimization.
By comparing actual vehicle trajectories with those obtained by minimizing travel distance (distance-minimizing trajectories, DMTs; see Sec.~\ref{sec:dataset} for time assignment) between the same OD pairs, we identify a systematic difference between the two.
At the beginning and end of trips, vehicles tend to take a distance greater than the DMTs, lose their directional memory quickly, and travel at lower speeds than during the middle part of trips.
In contrast, during the middle part of trips, vehicles tend to cover a distance close to the DMTs, maintain their directional memory, and travel at higher speeds than at the beginning and end of trips.
To explain these patterns, we constructed a double-layered network which mimics the hierarchical structure of real-world road networks (e.g., highways and low-speed roads).
Our simple model reproduces the key empirical observations qualitatively, suggesting that the observed patterns in vehicle movement arise from vehicles utilizing the hierarchical structure of road networks to minimize travel time.

\section{Dataset} \label{sec:dataset}
 We use large-scale individual vehicle trajectory data collected from navigation devices in individual vehicles provided by the Korea Transport Institute (KOTI).
 The dataset is constructed by map-matching GPS points to the edges of the road network, and includes the entry and exit times for each road segment, along with the corresponding speeds.
 Approximately 60 million individual vehicle trajectories were recorded across Korea in 2019.
 For the sake of simplicity, we focus on Jeju---an island which is not connected to the mainland by bridges or tunnels with a population of roughly 700,000---during January 2019, and on Seoul, the capital of Korea and a megacity with a population of roughly 9,600,000, from January 1 to January 7, 2019. Figure~\ref{fig:dataset}(a) and~\ref{fig:dataset}(b) illustrate examples of vehicle trajectories, starting from the blue point to the red point, in Jeju and Seoul, respectively.
 
 To focus on typical vehicle trajectories, we preprocessed the dataset by removing outliers with excessively short or long travel times.
 As shown in Fig.~\ref{fig:dataset}(c), the travel time distribution appears approximately log-normal.
Accordingly, we retained only trajectories whose travel times fall within the interval $[\mu-2\sigma,\mu+2\sigma]$ in logarithmic space, where $\mu$ and $\sigma$ are the mean and standard deviation of the logarithm of travel time, respectively.
 The shaded areas in Fig.~\ref{fig:dataset}(c) indicate the dataset used for analysis: travel times range from \SIrange{183}{5832}{\second} in Jeju and from \SIrange{212}{4703}{\second} in Seoul.
 After filtering, we obtained 147,620 individual vehicle trajectories in Jeju, and 122,652 trajectories in Seoul.

For each actual trajectory, we construct a DMT on the road network between its OD pair, which serve as a reference trajectory obtained solely by minimizing travel distance.
The blue line in Fig.~\ref{fig:sample_trajectory} shows one such DMT, while the red line represents the actual vehicle trajectory for the same OD pair.
To analyze not only the spatial but also the dynamical differences between actual trajectories and DMTs, we assign travel time to the road segments of each DMT by assuming constant-speed motion within each road segment.

Specifically, for each road segment $e$, we define its representative speed $v_\mathrm{seg}(e)$ as the mean speed of all vehicles that passed through $e$ in 2019.
We then assume that a vehicle travels road segment $e$ of the DMT at speed $v_\mathrm{seg}(e)$.
We adjust outliers in distribution of $v_\mathrm{seg}(e)$ across road segments in a road network, which fall outside the 5th and 95th percentiles, to the closest boundary values.
We also replace missing values, including those for segments with fewer than 30 data points, with the median value of the $v_\mathrm{seg}(e)$ distribution.

\begin{figure*}
    \centering
    \includegraphics[width=\linewidth]{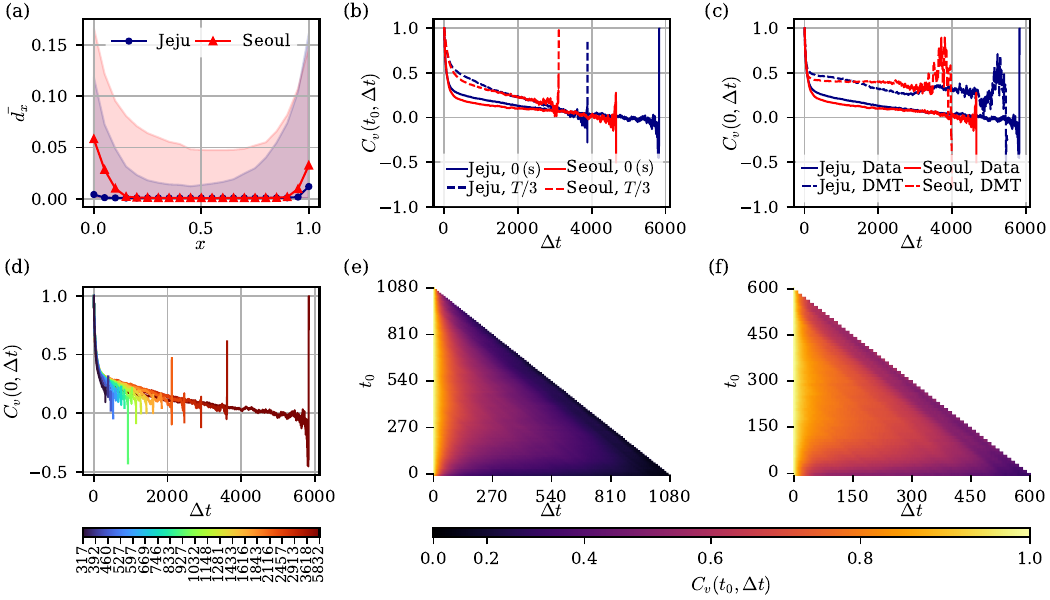}
    \caption{Spatial and dynamical measurements of real-world vehicle trajectories.
    (a) Median detour profiles for all trajectories in Jeju (blue) and Seoul (red).
    The shaded areas indicate the interquartile ranges (25th-75th percentile).
    (b) VDACF with reference times at the trip's beginning ($t_0=\SI{0}{\second}$, solid lines) and middle ($t_0=T/3$, dashed lines) for all trajectories, where $T$ denotes the total travel time of each trajectory.
    Blue lines represent Jeju, and red lines represent Seoul.
    (c) Comparison of the VDACF at $t_0=\SI{0}{\second}$ for vehicles in the empirical data (solid lines) and for vehicles minimizing travel distance between the same OD pairs (dashed lines).
    Blue lines represent Jeju, and red lines represent Seoul.
    (d) VDACF at $t_0=\SI{0}{\second}$ for trajectory groups in Jeju, classified by travel time, where each group contains an equal number of trajectories.
    The color of each line corresponds to the maximum travel time of the group, as indicated by the colorbar; for instance, the longest group includes trajectories with travel times between \SI{3618}{\second} and \SI{5832}{\second}.
    (e), (f) VDACF as a function of time lag ($\Delta t$) and reference time ($t_0$) for Jeju trajectories, to examine how VDACF changes when varying $t_0$, for a trajectory group with representative travel times: (e) actual trajectories (\SIrange{977}{1088}{\second}), (f) when vehicles minimize travel distance (\SIrange{538}{603}{\second}).
    For (e) and (f), colors are mapped using a power-law normalization ($\gamma = 1.5$) only to enhance visual contrast.
    }
    \label{fig:result}
\end{figure*}
 
\section{Results}
We compare actual vehicle trajectories with their DMTs for the same OD pairs in order to understand how vehicles move to minimize their cost function.

\subsection{Detour profile}
First, we measure the detour profile~\cite{mittal2024EfficientSelforganizationInformal} to quantify how much actual vehicle trajectories deviate from the minimum travel distance within sub-trajectories of a trip.
For a given trajectory, we define a sub-trajectory with a prescribed target length $\Delta L$ along the actual trajectory.
The variable $x \in \left[ 0,1 \right]$ denotes the normalized position of this sub-trajectory along the full trajectory.
In this notation, $x=0$ corresponds to the sub-trajectory starting from the origin, and $x=1$ corresponds to the sub-trajectory ending at the destination.
At each position $x$, the sub-trajectory has length $\ell_x$ measured along the actual trajectory.
We regard the two endpoints of this sub-trajectory as a local OD pair and compute the minimum travel distance $s_x$ between them on the road network.
The detour profile is defined as
\begin{equation}
    d_x=\frac{\ell_x-s_x}{\ell_x}.
\end{equation}
Thus, $d_x$ measures the fraction of the sub-trajectory length that exceeds the minimum travel distance between its two endpoints.
A value of $d_x=0$ indicates that the vehicle covers the minimum travel distance between the two endpoints of the sub-trajectory, whereas larger values indicate stronger local detours.
The limiting case $d_x=1$ corresponds to $s_x=0$, meaning that the vehicle returns to the same location after traveling a distance $\ell_x$.
In this work, following Ref.~\cite{mittal2024EfficientSelforganizationInformal}, we set the target length $\Delta L$ of the sub-trajectory to approximately one-third of the total trajectory length and investigate the median detour profile $\bar{d}_x$ over all trajectories.
Therefore, $\bar{d}_x$ quantifies how much longer vehicles travel than the minimum travel distance within sub-trajectories of their trips, revealing which parts of trips exhibit more prominent detours.
(For further details on this measure, see~\cite{mittal2024EfficientSelforganizationInformal}.)

Figure~\ref{fig:result}(a) illustrates the median detour profile results for Jeju (blue) and Seoul (red), with the interquartile range shaded in the corresponding colors.
As shown in Fig.~\ref{fig:result}(a), in both Jeju and Seoul, vehicles tend to travel much longer than the minimum distance at the beginning and end parts of their trips, while traveling nearly the minimum travel distance in the middle part of the trips.
These results suggest that for intervals in the middle parts of trips, the actual vehicle trajectories closely resemble the DMTs within the intervals.
In contrast, the results at $x\approx0$ and $x\approx1$ indicate that for intervals at the beginning and end of trips, actual trajectories differ significantly from the DMTs within the intervals.

Remarkably, the detour profiles exhibit a consistent tendency in both Jeju and Seoul, although quantitative differences still exist.
We attribute these differences to distinct geometric constraints.
For instance, Jeju has a large mountain at the center of the island, while Seoul is crossed by a large river flowing from east to west.
Considering that vehicle trajectories are influenced by the geometric structure of the road network (e.g., network dimension, road density, and interconnection with other cities)~\cite{boguna2009NavigabilityComplexNetworks, youn2008PriceAnarchyTransportation, braunstein2003OptimalPathsDisordered}, the difference in $\bar{d}_x$ values between Jeju and Seoul appears to stem from such geometric constraints.
Nevertheless, the consistent tendency of the detour profiles indicates that the observed patterns emerge from the intrinsic structural properties shared by road networks, rather than from regional characteristics.
Indeed, if vehicle trajectories are determined by the nature of road networks, this pattern should manifest not only in the detour profile, which is a spatial measure, but also in dynamic measures.

\subsection{Velocity direction autocorrelation function}
To investigate how the observed detour patterns manifest dynamically, we calculate the velocity direction autocorrelation function (VDACF) for all  trajectories~\cite{dalessandro2018CollectiveRegulationCell, rehfeldt2023RandomWalkersToolbox}, denoted as $C_v(t_0, \Delta t)$.
The VDACF is defined as
\begin{equation}
C_v(t_0,\Delta t) = \left\langle \frac{\vec{v} \left( t_0 \right)}{{\left|\vec{v} \left( t_0\right)\right|}} \cdot \frac{\vec{v} \left( t_0+\Delta t\right)}{\left| \vec{v}\left( t_0+\Delta t\right)\right|}\right\rangle,
\end{equation}
where $\vec{v}\left( t\right)$ is the velocity at time $t$, $t_0$ is a reference time point, and $\Delta t$ is the time lag.
Here, $\langle \cdots \rangle$ denotes the ensemble average across all trajectories.
$C_v$ quantifies the persistence of directional memory over time, capturing how consistently vehicles maintain their direction of motion as time progresses.
In other words, we can examine how the persistence of vehicle direction decays over a time window $\Delta t$ from the reference time $t_0$.
To calculate $C_v$ of actual trajectories, we assume that the vehicle travels at a constant speed along each road segment of its trajectory.
For each vehicle, the travel time on a given segment is obtained by measuring the difference between its entry and exit times on that segment.
We then divide each trajectory into 1-second interval points based on these segment travel times and calculate the VDACF using the velocity vectors at these 1-second intervals.

To identify that the disparity observed in $\bar{d}_x$ manifests in $C_v$, we compute the VDACF at two distinct reference points: $t_0=\SI{0}{\second}$, representing the beginning part of a trip, and $t_0=T/3$, where $T$ denotes the total travel time of each trajectory, representing the middle part of the trip.
As shown in Fig.~\ref{fig:result}(b), our analysis reveals that the VDACF with $t_0 = T/3$ (dashed lines) decays significantly slower than the VDACF with $t_0 =\SI{0}{\second}$ (solid lines).
This result suggests that vehicles tend to maintain their direction longer in the middle part of trips.
Conversely, vehicles change their direction frequently in the beginning part of the trip, as shown by a more rapid decay of the VDACF.
Furthermore, this disparity of directional memory is highly consistent with the detour profile results, which showed greater detours at the beginning and reduced detours in the middle.
Moreover, we compare the $C_v(t_0,\Delta t)$ at $t_0=\SI{0}{\second}$ for vehicles in empirical data with that for vehicles minimizing travel distance between the same OD pairs in both Jeju and Seoul, which is represented in Fig.~\ref{fig:result}(c).
To compute VDACF of vehicles minimizing travel distance, we utilize the mean speed of each road segment $v_\mathrm{seg}(e)$ as described in Sec.~\ref{sec:dataset}.
We assume that the velocity of vehicle following DMT is given by these average speeds.
A primary observation is that $C_v(0, \Delta t)$ of the actual vehicles (solid lines) decays faster than that of vehicles minimizing travel distance (dashed lines) for both Jeju and Seoul.

It is, however, challenging to analyze the VDACF characteristics at the end parts of trips, since trips of different length are averaged into $C_v(0, \Delta t)$, resulting in the loss of fine details.
To address this issue and ensure robust analysis, we divide the trajectories into 20 distinct bins based on their total travel times, ordered from the shortest to the longest trips, with each bin containing an equal number of trajectories.
Remarkably, as shown in Fig.~\ref{fig:result}(d), the results show consistent and clear patterns after binning: the beginning, middle, and end parts of the trips.
Specifically, the VDACF exhibits a rapid decay at the beginning parts of the trips, followed by a much slower rate of decay during the middle parts, and then decays rapidly again at the end parts of the trips.
This tripartite dynamical signature is observed consistently across all 20 travel time bins, highlighting its generality regardless of total travel time.

\begin{figure*}
    \centering
    \includegraphics[width=\textwidth]{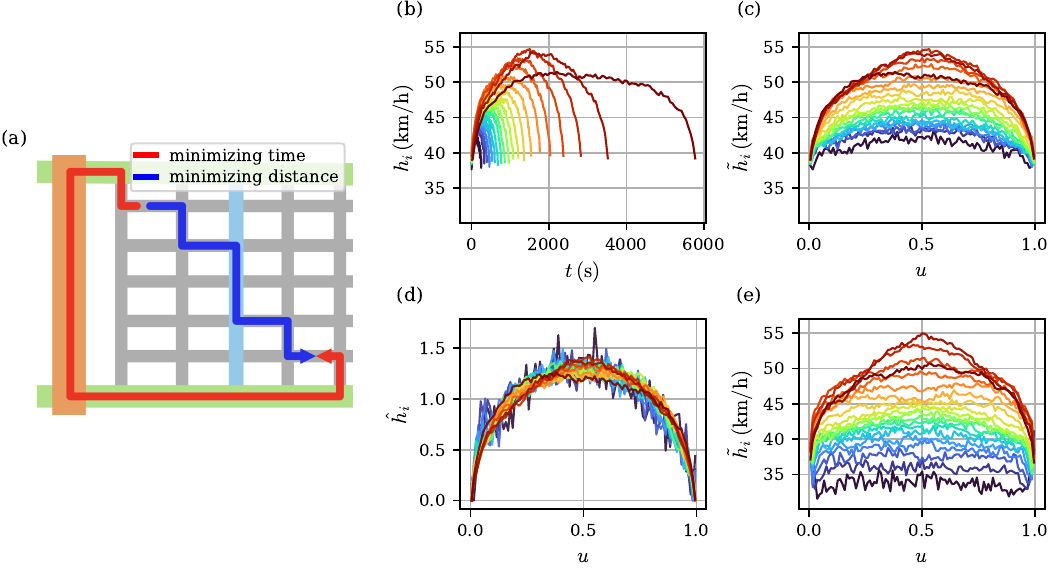}
    \caption{Hierarchical patterns in vehicle trajectories.
    (a) Schematic of trajectories that minimize travel time (red line) and travel distance (blue line).
    When minimizing travel time on the hierarchical network, a vehicle utilizes high-level roads (orange and green lines) to reduce travel time, despite traveling a longer distance than when it minimizes travel distance.
    (b) Temporal changes in road hierarchy for actual trajectories in Jeju.
    The hierarchy is quantified by the representative speed $v_{\mathrm{seg}}(e)$ of each road segment $e$, defined as the mean speed of vehicles passing through that segment.
    For trajectory group $i$, $h_i(t)$ denotes the ensemble averaged $v_{\mathrm{seg}}$ at time $t$.
    Each color represents the same travel time group as in Fig.~\ref{fig:result}(d).
    (c) Same as (b), but with time linearly normalized to $u \in [0,1]$, where $u=0$ and $u=1$ correspond to the origin and destination, respectively; the resulting profile is denoted by $\tilde{h}_i(u)$.
    (d) Rescaled profile $\hat{h}_i(u)$, showing nearly identical behavior across groups.
    (e) Same analysis as in (c), but conducted for the DMTs between the same OD pairs, grouped into 20 bins by their travel times.}
    \label{fig:Hierarchy}
\end{figure*}

To investigate the change in tendency when varying the reference time $t_0$, we select actual trajectories whose travel times fall within the 5th-percentile range centered around the median in Jeju (\SIrange{977}{1088}{\second}) and compute the VDACF across all possible reference times $t_0$ and time window $\Delta t$, as illustrated in Fig.~\ref{fig:result}(e).
(We also calculated the VDACF with the time-reversal protocol for the end part of the trip and obtained a similar result.
See the Appendix~\ref{appendix:time-reversal}.)
As shown in Fig.~\ref{fig:result}(e), the VDACF decays rapidly at the beginning and end parts of trips.
In contrast, within the middle part of the trips, the $C_v$ remains significantly high over a long time lag $\Delta t$.
These findings indicate a common property of vehicular movement, revealed by both static and dynamic measures, which is the distinguishability into beginning, middle, and end parts of trips.

Similarly, to compare how the VDACF of actual vehicles differs from that of vehicles minimizing travel distance between the same OD pairs when varying the reference time $t_0$, we apply the same binning procedure to the DMTs based on their total travel times.
We then select DMTs whose travel times fall within the 5th-percentile range around the median (\SIrange{538}{603}{\second}) and compute the VDACF across all possible reference times $t_0$ and time windows $\Delta t$; the results are shown in Fig.~\ref{fig:result}(f).
When vehicles minimize their travel distance, the $C_v$ remains remarkably higher across a broad range of $t_0$ and $\Delta t$ compared with that of the actual vehicles.
Although the VDACF of vehicles minimizing travel distance may appear to decay rapidly at the beginning and end of trips (i.e., $t_0\approx0$ and $t_0\approx1$), it actually remains at relatively high values.
Moreover, in the middle parts of trips, the VDACF of vehicles minimizing travel distance exhibits strong directional persistence for significantly longer than that observed from actual vehicles.

\subsection{Road hierarchy}
We hypothesize that the observed static and dynamic patterns in actual vehicle trajectories originate from the hierarchical structure of road networks.
That is, since a broad range of low-level roads (e.g., low-speed roads) are interconnected by high-level roads (e.g., highways) at high speeds, vehicles minimizing travel costs tend to exhibit specific patterns at the beginning, middle, and end of their trajectories.
Under these conditions, if a vehicle minimizes the travel time, a vehicle routes to utilize high-level roads whenever the time savings by using high-level roads are significant enough, even if it means taking a detour.
This results in a high detour profile and the rapid decay of VDACF when accessing such roads.
In contrast, when a vehicle minimizes its travel distance, it may avoid high-level roads because, even though these roads allow for faster travel over long distances, using them may require a longer detour.
Consequently, the tendency to maintain direction towards the destination is stronger than the tendency to utilize high-level roads.
As a result, the VDACF of vehicles minimizing travel distance decays more slowly than that of vehicles minimizing travel time.

Figure~\ref{fig:Hierarchy}(a) presents a schematic diagram illustrating the trajectories generated when a vehicle minimizes its travel time (red line) and when it minimizes its travel distance (blue line).
This suggests that a trajectory when a vehicle minimizes its travel time consists of three distinct phases.
Phase 1 involves accessing high-level roads (illustrated as the orange and green lines) at the beginning part of the trip.
In Phase 2, vehicles rapidly travel toward the vicinity of the destination by utilizing these high-level roads.
Finally, Phase 3 involves exiting the high-level roads and reaching the final destination through low-level roads.
Since the time advantage gained from using high-level roads is greater than the additional distance cost caused by detours on low-level roads, we expect that this low-high-low road usage pattern will be observed in hierarchical road networks.
This interpretation is consistent with the patterns observed in actual vehicle trajectories.
Conversely, when vehicles minimize their travel distance, the hierarchical structure of road networks has much less influence on the vehicle movement.
Thus, this low-high-low road usage pattern would not emerge.
From this perspective, the observed high detour and rapid VDACF decay at the beginning of trips can be interpreted as the process of accessing high-level roads and, symmetrically, those at the end as the process of exiting them.

To support our hypothesis, we examine the temporal changes in road hierarchy during vehicle movement.
For each road segment $e$, we assign its representative speed $v_{\mathrm{seg}}(e)$, defined as the mean speed of vehicles passing through that segment, as described in Sec.~\ref{sec:dataset}.
We then group trajectories by total travel time, as in the VDACF analysis.
For each group $i$, we compute the hierarchy profile $h_i(t)$, defined as the ensemble average of $v_\mathrm{seg}$ at time $t$ over trajectories in group $i$.
Figure~\ref{fig:Hierarchy}(b) illustrates that $h_i(t)$ exhibits an inverted-U shaped pattern of road hierarchy usage in the empirical dataset for different travel time groups.
To simplify comparison across groups, we linearly normalize time so that the origin and destination correspond to $u=0$ and $u=1$, respectively, and denote the resulting profile by $\tilde{h}_i(u)$, as shown in Fig.~\ref{fig:Hierarchy}(c).
This inverted-U shaped hierarchy usage is not only observed in road networks but also in other human-utilized networks~\cite{boguna2009NavigabilityComplexNetworks, gulyas2020RoleDetoursIndividual}.
Furthermore, to compare the profile shape across groups, we investigate the rescaled hierarchy profile $\hat{h}_i(u)$, defined as
\begin{equation}
    \hat{h}_i(u)=\frac{\tilde{h}_{i}(u)-\tilde{h}_{i,\min}}{\tilde{h}_{i,\mathrm{avg}}-\tilde{h}_{i,\min}},
\end{equation}
where $\tilde{h}_{i, \min}$ and $\tilde{h}_{i,\mathrm{avg}}$ are the minimum and the average of $\tilde{h}_{i}(u)$ over $u$, respectively.
As illustrated in Fig.~\ref{fig:Hierarchy}(d), the temporal changes of hierarchy across different travel times exhibit nearly identical behavior.
In contrast, as shown in Fig.~\ref{fig:Hierarchy}(e), when vehicles minimize travel distance, only relatively minor changes are observed across groups except for those with long travel times.
While these results generally support our hypothesis, Fig.~\ref{fig:Hierarchy}(e) also shows that vehicles minimizing travel distance exhibit an inverted-U shaped pattern in road hierarchy usage as travel time increases.
This indicates that vehicle trajectories are influenced by the structural constraints of road networks.
In particular, as the distance between OD increases, the DMT necessarily includes high-level roads, because continuous connections of low-level roads do not span long distances.

\section{Model}

\begin{figure*}
    \centering
    \includegraphics[width=\textwidth]{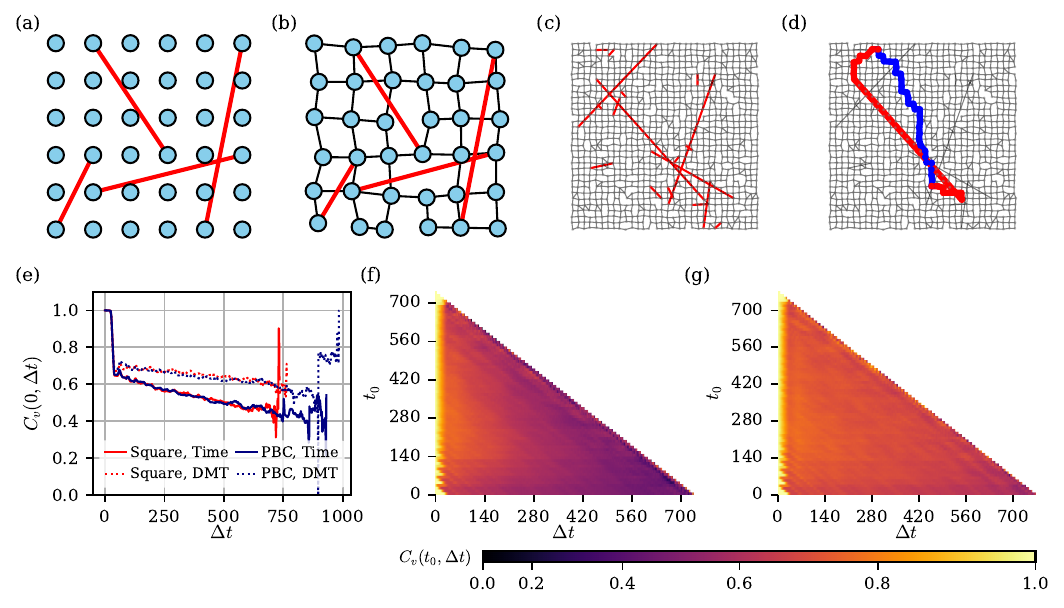}
    \caption{Model for double-layered road networks and vehicle movement on the model networks
    (a) Schematic illustrating the placement of high–level roads (red lines) on a grid of nodes.
    The probability of adding a high-level road with Manhattan distance $L$ follows the power law $L^{-\alpha}$ and the high-level road addition process stops once the total sum of high-level roads' Manhattan distance exceeds the budget $C$.
    (b) Schematic illustrating the formation of the low-level road network.
    Node positions are perturbed, and adjacent nodes are connected by low-level roads (black lines) to simulate imperfections of road networks.
    (c) An example of a network generated with parameters ($N=32,\ C=5Nd,\  \alpha =2,\ \tau=0.5,\ r=1.2d)$, where high-level roads are depicted in red.
    (d) Sample trajectories of a vehicle minimizing its travel time (red line) and travel distance (blue line) for the same OD pair on the network shown in (c).
    (e) Comparison of the VDACF at $t_0=\SI{0}{\second}$ for vehicles minimizing travel time (solid lines) and vehicles minimizing travel distance between the same OD pairs (dotted lines) on the square lattice-like networks (red) and the networks with PBC (blue).
    (f), (g) Full VDACF profiles computed from the central 1,000 trajectories per network in terms of travel time, $C_v(t_0, \Delta t)$, on the square lattice-like networks for (f) vehicles minimizing travel time and (g) vehicles minimizing travel distance.
    For (f) and (g), colors are mapped using a power-law normalization ($\gamma = 1.5$) only to enhance visual contrast.}
    \label{fig:square_model}
\end{figure*}

To verify our hypothesis and analyze the influence of the hierarchical structure of road networks, we construct a simple double-layered network model.
We consider a modified version of the model introduced in Ref.~\cite{Li2010DesignPrinciplesOptimal}.
Our model consists of distinct high-level and low-level roads, mimicking real-world road hierarchies.
To represent the distinct urban structures of Seoul and Jeju, we implement two different types of network topologies: a square lattice-like network for the grid-like nature of Seoul, and a network with periodic boundary conditions (PBC) to represent the island geography of Jeju.

First, we construct a square lattice network composed of $N \times N$ nodes with an internode distance of $d$, and then we add high-level roads whose segments are longer and whose speeds are faster.
The highway addition process stops once the total sum of their Manhattan distances exceeds the construction budget $C$; the Manhattan distance is used here because it fully captures the routing cost of the underlying square lattice network.
The probability of adding a highway with Manhattan distance $L$ follows the power law $P(L) \propto L^{-\alpha}$ [Fig.~\ref{fig:square_model}(a)].
The relative speed advantage of high-level roads is controlled by a time efficiency parameter, $\tau$ (with $0<\tau<1$), such that the speed on high-level roads, $v_h$, is set relative to the speed on low-level roads, $v_l$, by the relation $v_h=v_l/\tau$.
Additionally, to implement road network defects and to make the shortest path between specific nodes unique, we introduce fluctuations to the nodes' positions within a radius $d/5$, and rewire every edge for which the internode distance is less than a threshold $r$ [Fig.~\ref{fig:square_model}(b)].
A realization of our model network is illustrated in Fig.~\ref{fig:square_model}(c), where the highways are depicted as red lines.
Here, we set $N=32,\ C=5Nd,\ \alpha = 2,\ \tau=0.5,\ r=1.2d$ for the analysis of our network model.
To create the network with PBC, we double the number of nodes in the $x$-direction and halve it in the $y$-direction, and then connect the nodes at the opposing ends of the $x$-axis to implement periodicity.
For the analysis, we generate 100 networks under each condition and produce 3,000 trajectories for randomly selected OD pairs on each network.
Here, we assume that $d \approx \SI{1}{\kilo\meter}$ and that vehicles travel at \SI[per-mode=symbol]{30}{\kilo\meter\per\hour} on low-level roads.

To examine whether the simple model can reproduce the patterns observed in empirical data, we analyze the VDACF of vehicles traveling on our model networks.
Figure~\ref{fig:square_model}(d) illustrates sample trajectories generated on the model networks.
For the same OD, the red line represents a trajectory that minimizes travel time, whereas the blue line represents one that minimizes travel distance.
This visualization clearly shows that characteristic behaviors observed in the empirical data---such as frequent direction changes at the beginning and end of trips---are successfully reproduced by our model.
Figure~\ref{fig:square_model}(e) demonstrates that the VDACF of vehicles minimizing travel time decays more rapidly than that of vehicles minimizing travel distance on our hierarchical model networks, consistent with our main results.
In Fig.~\ref{fig:square_model}(f) and (g), the VDACF was computed using the central 1,000 trajectories in terms of travel time from each network.
As shown in Fig.~\ref{fig:square_model}(f), the VDACF of vehicles minimizing travel time on our model networks decays quickly at small and large $t_0$.
In stark contrast, as shown in Fig.~\ref{fig:square_model}(g), when vehicles minimize travel distance, the VDACF exhibits persistence across the entire range of $t_0$.
These VDACF patterns qualitatively resemble those observed for actual vehicles and for vehicles minimizing travel distance between the same OD pairs, as illustrated in Fig.~\ref{fig:result}(d) and (f).
Moreover, the results for the model with PBC are similar to those for the square lattice-like model and are illustrated in the Appendix~\ref{appendix:PBC}.

\begin{figure}
    \centering
    \includegraphics[width=0.8\linewidth]{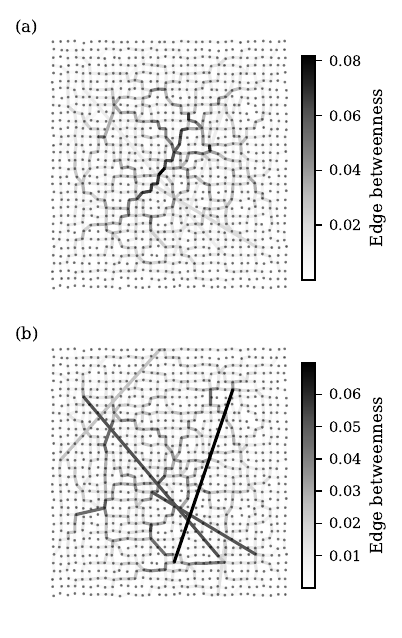}
    \caption{Visualization of edge betweenness centrality under two different weighting schemes.
    (a) In the distance-weighted network, the centrality is distributed across the low-level roads, even though the high-level roads connect distant regions.
    (b) In the time-weighted network, the centrality becomes highly concentrated on the high-level roads.}
    \label{fig:EB}
\end{figure}

To verify the influence of high-level roads on road networks, we investigate the edge betweenness centrality for networks weighted by either travel distance or travel time.
Since edge betweenness centrality quantifies how frequently a road segment is utilized in all shortest paths, it highlights the essential road segments which affect the connection of entire networks~\cite{kwon2023GlobalEfficiencyNetwork, akbarzadeh2019RoleTravelDemand}.
As shown in Fig.~\ref{fig:EB}(a), when we calculate the centrality based on the distance, high-level roads do not play an important role, even though they connect distant regions.
In contrast, as illustrated in Fig.~\ref{fig:EB}(b), when we calculate the centrality based on travel time, these roads exhibit significantly higher values of edge betweenness centrality.
Based on these results, we can conclude that the travel-time-based hierarchical structure of road networks leads to the utilization of high-level roads.
Specifically, high-level roads, which cover long distances at high speeds, significantly reduce travel time, concentrating vehicles onto these roads.
Therefore, when vehicles minimize travel time, the process of accessing and exiting these high-level roads causes the rapid decay of directional memory at the beginning and end parts of trips.

\section{Discussion}
In summary, we addressed questions of how vehicles actually move in terms of speed and direction on real-world road networks to minimize their costs and how the underlying network structure affects their movements.
By analyzing large-scale individual vehicle trajectory datasets and comparing them with the DMTs for the same OD pairs, we revealed the characteristics of individual vehicle trajectories.
Using the detour profile and VDACF, we observed that vehicles typically take more detours and lose their directional memory faster at the beginning and end of their trips compared to the middle.
In contrast, they tend to travel with minimal detours and maintain high speeds in the middle parts of their trips.
We hypothesize that these patterns emerge from the hierarchical structure of road networks based on the observation of inverted-U shaped hierarchy profile in the real-world trajectory data.
To verify our hypothesis, we constructed simple double-layered road networks and produced trajectories for vehicles minimizing either their travel time or distance on these networks.
This simple model qualitatively reproduced the VDACF patterns observed in actual vehicle trajectories when vehicles minimize their travel time, utilizing high-level roads.
Furthermore, our edge betweenness centrality analysis revealed that the high-level roads attract vehicles when they minimize travel time, whereas this is not the case when minimizing travel distance.

Our results suggest that vehicle trajectories exhibit a consistent regularity that can be divided into distinct phases: the beginning, middle, and end of trips, regardless of travel time or region.
These phases emerge when vehicles move on hierarchical road networks that allow them to cover long distances in a short time with fewer directional changes while traveling on high-level roads, and to access high-level roads via low-level roads with more directional changes, thereby minimizing total travel time.
Consequently, to enter and exit these high-level roads, vehicles take routes that are significantly different from the DMT at the beginning and end of trips.
Through this process, vehicle trajectories exhibit an inverted U-shaped hierarchical road usage pattern.
Interestingly, such an inverted U-shaped hierarchical pattern reflects a consistency in human behavior across various networks~\cite{boguna2009NavigabilityComplexNetworks, gulyas2020RoleDetoursIndividual}.
Moreover, combined with findings of Ref.~\cite{wang2012UnderstandingRoadUsage}, our results suggest that maintaining high flow on high-level roads is more efficient than improving numerous low-level roads~\cite{kwon2023GlobalEfficiencyNetwork}.
This is because high-level roads, although they constitute only a small fraction of the road network, significantly affect the entire system.

From the perspective of cost minimization, this suggests that even a small fraction of high-level roads can substantially reduce the effective travel cost over the network.
This large cost reduction geometrically suggests the emergence of a `small-world' network structure, which means that these high-level roads effectively shorten path lengths between OD pairs~\cite{watts1998CollectiveDynamicsSmallworld, latora2001efficientbehaviour, goodrich2022modelingsmallworld}.
Our results suggest that this effect is particularly pronounced when travel time is taken as the travel cost.
In terms of travel time, high-level roads effectively shrink the path length between distant points, thereby emphasizing local network structures near their entry and exit points because of the expected concentration of traffic demand~\cite{liu2022digitaltwinhighway}. 
This geometric distortion caused by hierarchical structures provides an important perspective for understanding the nature of transportation systems, as many such systems inherently possess hierarchical properties to improve efficiency.

Although our results reveal the regularity of individual vehicle trajectories and the fundamental elements that affect them, there are some limitations.
First, our research does not explain the deviation of actual vehicle trajectories from the trajectories that minimize travel time or the existence of multiple paths between the same OD pairs.
However, we speculate that the fundamental routing principle is still based on optimizing travel time (i.e., Wardrop's first principle), and such deviations emerge from dynamic changes in flow (due to congestion, commuting patterns, etc.) and individual-level fluctuations~\cite{zhang2009IllusionMotionVariation, levinson2013PortfolioTheoryRoute, parthasarathi2013NetworkStructureTravel, lima2016UnderstandingIndividualRouting}.
Second, in this work, since our model is simulated with fixed parameters to focus on explaining trajectory patterns, we do not analyze the effects of each model parameter.
We leave a detailed parameter analysis of the model, such as identifying the effects of high-level road length, speed, and budget for future work.

\begin{acknowledgments}
We thank J.D. Noh for helpful discussions.
This research was supported by Basic Science Research Program through the National Research Foundation of Korea (NRF) grant funded by the Ministry of Education (Grant No. RS-2024-00346558) and the Ministry of Science and ICT (Grant No. RS-2025-00515556).
The authors also acknowledge the Urban Big data and AI Institute of the University of Seoul supercomputing resources (https://ubai.uos.ac.kr) made available for conducting the research reported in this paper.
\end{acknowledgments}

\appendix

\section{VDACF for time-reversal trajectory} \label{appendix:time-reversal}
\begin{figure}[h!]
    \centering
    \includegraphics[width=\linewidth]{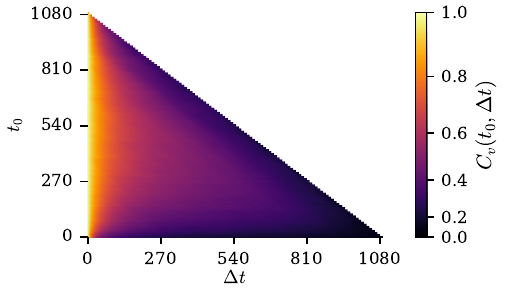}
    \caption{
    VDACF calculated with the time-reversal protocol for actual trajectories in Jeju. The reference time $t_0=\SI{0}{\second}$ corresponds to the arrival at the destination, and $\Delta t$ represents the backward time lag.
    The rapid decay near $t_0=\SI{0}{\second}$ indicates that vehicles lose directional memory quickly at the end part of trips.
    Colors are mapped using a power-law normalization ($\gamma=1.5$) only to enhance visual contrast.
    }
    \label{fig:time-reversal}
\end{figure}

As total travel times vary across individual trajectories, averaging the VDACF forward from the origin $(t_0=\SI{0}{\second})$ obscures the dynamic characteristics near the destination.
To address this, we apply a time-reversal protocol, setting the arrival time as the reference time and calculating the VDACF backward.
As shown in Fig.~\ref{fig:time-reversal}, the VDACF for time-reversal trajectories exhibits a rapid decay near the destination, confirming that vehicles also lose directional memory quickly during the end part of trips.

\section{Periodic boundary conditions} \label{appendix:PBC}

\begin{figure}[h!]
    \centering
    \includegraphics[width=\linewidth]{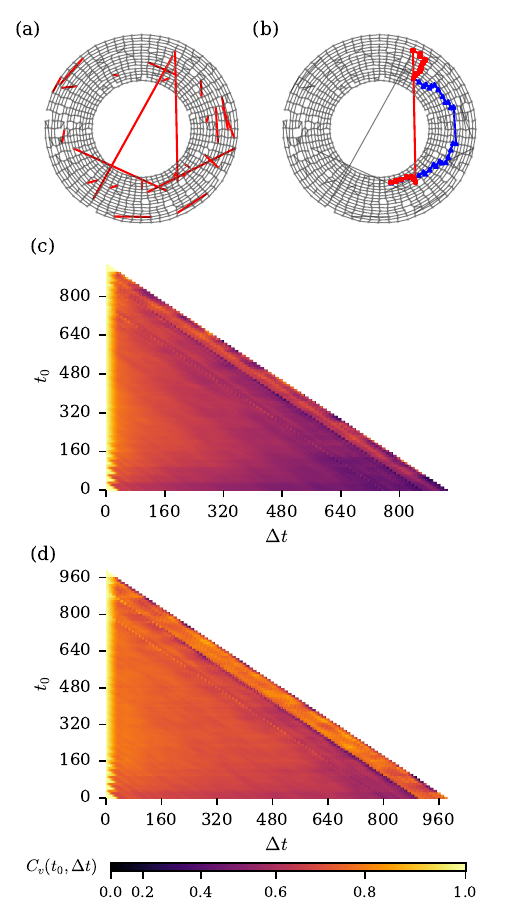}
    \caption{Modeling hierarchical road networks with PBC.
    (a) An example of a network generated with PBC, where high-level roads are depicted in red.
    (b) Sample trajectories of a vehicle minimizing its travel time (red line) and distance (blue line) for the same OD pair on the network shown in (a).
    (c), (d) Full VDACF profiles, $C_v(t_0, \Delta t)$ of (c) vehicles minimizing travel time and (d) vehicles minimizing travel distance.
    For (c) and (d), colors are mapped using a power-law normalization ($\gamma = 1.5$) only to enhance visual contrast.}
    \label{fig:pbc}
\end{figure}

We also simulate vehicle routing on a hierarchical network with PBC, mimicking the geography of an isolated island like Jeju.
We double the nodes in the $x$-direction and halve it in the $y$-direction, and connect opposing ends of the $x$-axis.
As shown in Fig.~\ref{fig:pbc}(c), the VDACF of vehicles minimizing travel time on this periodic network exhibits the same characteristic fast-slow-fast decay pattern observed in the square lattice-like model.

\FloatBarrier
\bibliography{csns2026}

\end{document}